# Transistors with Chemically Synthesized Layered Semiconductor WS$_2$ Exhibiting 10$^5$ Room Temperature Modulation and Ambipolar Behavior

Wan Sik Hwang[1,a)], Maja Remskar[2], Rusen Yan[1], Vladimir Protasenko[1], Kristof Tahy[1], Soo Doo Chae[1], Pei Zhao[1], Aniruddha Konar[1], Huili (Grace) Xing[1], Alan Seabaugh[1], and Debdeep Jena[1,b)]

[1]Department of Electrical Engineering, University of Notre Dame, Notre Dame, IN 46556, USA
[2]Solid State Physics department, Jozef Stefan Institute, Jamova 39, SI-1000 Ljubljana, Slovenia

a) whwang1@nd.edu
b) djena@nd.edu

**ABSTRACT**

We report the realization of field-effect transistors (FETs) made with chemically-synthesized layered two dimensional (2D) crystal semiconductor WS$_2$. The 2D Schottky-barrier FETs demonstrate ambipolar behavior and a high (~10$^5$x) on/off current ratio at room temperature with current saturation. The behavior is attributed to the presence of an energy bandgap in the 2D crystal material. The FETs show clear photo response to visible light. The promising electronic and optical characteristics of the devices combined with the layered 2D crystal flexibility make WS$_2$ attractive for future electronic and optical devices.

The continued success of modern electronic devices has been driven by scaling down of the active area such as channel length and gate dielectric thickness. As device sizes continue to shrink, short channel effects lead to poor threshold control [1]. The origin of such effects is linked to the 3-dimensional (3D) nature of semiconductors used to make the transistors. Alternatively, silicon on insulator (SOI) technology was introduced in order to reduce device degradation due to short-channel effects by rendering the active semiconducting material more two-dimensional [2]. The recently discovered material graphene [3] on the other hand is a true 2D material with a single atomic sheet of graphite that is highly attractive for scaling from electrostatics viewpoint, but has a zero energy band gap. It has sparked intense research for electronic device applications due to its exceptionally high mobility (excess of 15,000cm$^2$/Vs) at room temperature and high Fermi velocity [4], yet the absence of an energy gap prevents it from traditional electronic switching since it cannot be switched off. Efforts to overcome this problem involve opening of bandgaps by quantum confinement in graphene nanoribbons (GNRs) [5]. The size of the energy gap is inversely proportional to the width of the GNR [7]. GNR field-effect transistors (FETs) using exfoliated graphene [6, 7], epitaxial graphene [8], or chemical vapor deposition (CVD) graphene [9] close to ~ 10 nm widths still exhibit low $I_{ON}/I_{OFF}$ ratios at room temperature due to the small gaps; the effect of edge disorder and absence of current saturation currently are proving major challenges. Novel device concepts such as interlayer tunneling FETs may help graphene based FETs achieve better switching in the near future [10, 11], but the zero-gap nature of 2D graphene remains a hindrance for traditional planar FET realization with this 2D crystal.

Spurred by the knowledge of isolation of graphene, other 2D transition-metal dichalcogenide materials in the form of MX$_2$ (where M=transition metal such as Mo, W, Ti, Nb, etc. and X=S, Se, or Te) have drawn considerable attention. The multilayer forms of these materials, like graphite, are traditionally used as lubricants or intercalated batteries due to their layered structure, and are thus chemically synthesized in large volumes. The $MX_2$ family material consists of one or more sets of triple layers with one M and two X in a sandwich structure (X-M-X layers). Atoms within each layer are strongly held together by covalent-ion mixed bonds, while interlayer van der Waals forces are weak. Substantial prior works in 2D materials have concentrated on optical and material properties [12~14]. FETs using $MoS_2$ and $WSe_2$ have been demonstrated with substantial gate modulation [15, 16]. A recent calculation showed that single layer $WS_2$ has the potential to outperform Si and other 2D crystals in FET-type applications due to its favorable bandstructure [17]. No prior device results have been reported for $WS_2$ for logic or optical devices. In this letter, we report the first fabrication and demonstration of 2D $WS_2$ FETs and also explore the effects of photoexcitation on the transistor characteristics.

$WS_2$ flakes were grown by the iodine transport method from previously synthesized $WS_2$ (0.6 g) at 1060 K in evacuated silica ampoule at pressure of $10^{-3}$ Pa, and with temperature gradient of 6.8 K/cm. The volume concentration of iodine was 11 mg/cm$^3$. After 21 days of growth the silica ampoule was slowly cooled to room temperature with a controlled cooling rate of 30 °C / hour. The $WS_2$ flakes were then dispersed by sonication in isopropyl alcohol (IPA) and deposited on 30 nm atomic-layer-deposited (ALD) $Al_2O_3$/*p*-Si substrates at 170 °C to dry the solution. A schematic cross-sectional image of the back-gated $WS_2$ device is shown in Fig. 1(a). The source and drain contacts are defined by electron beam lithography (EBL) using Ti/Au (5/100 nm) contacts. An atomic force microscope (AFM) image of the $WS_2$ device with *L/W* = 2.5/2 μm is shown in Fig 1(b). The cross-sectional height of $WS_2$ layers is ~13nm which is equivalent to 18~20 single layers (interlayer spacing of $WS_2$ is in the range of 0.65~0.70nm [18]). The Raman spectra ($\lambda_{exc}$ = 488 nm) of the $WS_2$ region as shown in Fig 2(d) exhibits two peaks: one in the $E_{2g}^1$ range for in-plane vibrations at ~356 $cm^{-1}$ and the other in the $A_{1g}$ range for out-of-plane vibrations at ~421 $cm^{-1}$ [19]. The laser spot is aligned to the channel region with spot size of 0.5-1 μm and laser power of ~1.5mW. The 2D Raman signal is fit to two single Lorentzian models, revealing that the chemical vapor deposited (CVD) 2D $WS_2$ retains the single crystal properties of $WS_2$ with unnoticeable structural modifications.

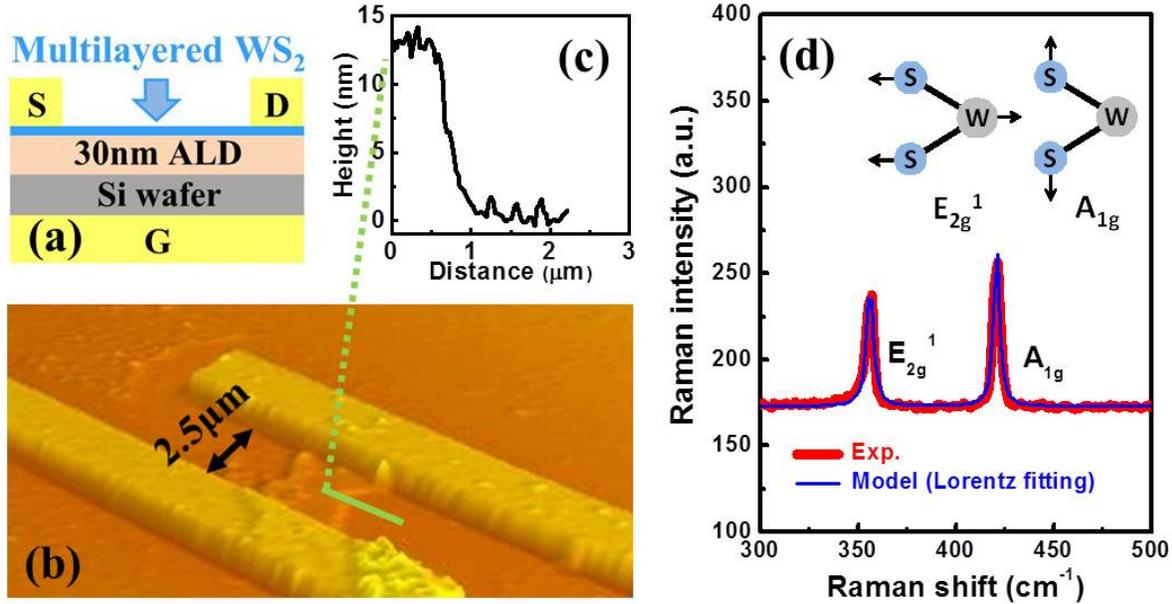

**FIG**. 1. (a) Schematic cross-sectional image of a$WS_2$ transistor with multilayered $WS_2$. (b) Atomic force microscope (AFM) image of the device on top of 30 nm $Al_2O_3$ layer with Ti/Au contacts. (c) Cross-sectional height of $WS_2$ layer indicated at green line of Fig. 1(b) by AFM. (d) Raman spectra (λ = 488 nm) of the multilayered $WS_2$ with laser power of 1.5mW. The insert

sketch shows the two primary vibrational modes of WS$_2$ leading to the two peaks in the Raman spectrum.

Figure 2(a) shows the measured drain current $I_D$ versus the gate-source voltage $V_{GS}$ at room temperature for a multilayer WS$_2$ device at two drain biases. The gate modulation is ~10$^5$x for $V_{DS}$=1 V, and ~10$^4$ for $V_{DS}$=5 V. The gate leakage current is much lower (~few pA) than the drain current and is not shown for clarity. The device shows clear ambipolar behavior indicating accumulation of electrons (n-type conductivity) for positive $V_{GS}$ and of holes (p-type conductivity) for negative $V_{GS}$ regions. Thus, electron and hole carriers are preferentially injected depending on the gate bias as illustrated in the insert of Fig. 2(a) as a consequence of Schottky barrier contacts. This is seen clearly in the family of $I_D$ - $V_{DS}$ curves in Fig 2(b). The high resistance of the Schottky barrier contacts also makes it difficult to extract important parameters like carrier motilities in a transparent manner from the field-effect behavior. It is expected that low-resistance ohmic contacts will be possible for WS$_2$ in the near future. Several methods are available to suppress one of the carriers to make the devices less ambipolar to decrease the $I_{off}$ and improve the modulation [20, 21]. Nevertheless, the ambipolar behavior can be advantageous for CMOS-like inverter applications [22], or for analog frequency multiplication purposes [23]. The energy band line-ups as shown in the insert of Fig. 2(b) indicates that Fermi level of the contact metal is aligned in the band gap of WS$_2$. The contact metal consists of Ti/Au (5/100 nm) and though the work function of Ti is 4.3 eV, if the thickness of metal (Ti in this work) is less than 5 nm, the net work function of the metal stack can be influenced by the 2$^{nd}$ metal (Au in this work) [24, 25]. The family of $I_D$ - $V_{DS}$ curves at various $V_{GS}$ in Fig 2(b) shows current saturation, a feature enabled by the substantial bandgap. The saturation of current is attributed to pinch-off in the channel at the drain side, similar to long-channel transistors. Both the high on/off current ratio (~$10^5$x) and current saturation over a wide voltage window are possible due to the presence of a bandgap, unlike graphene.

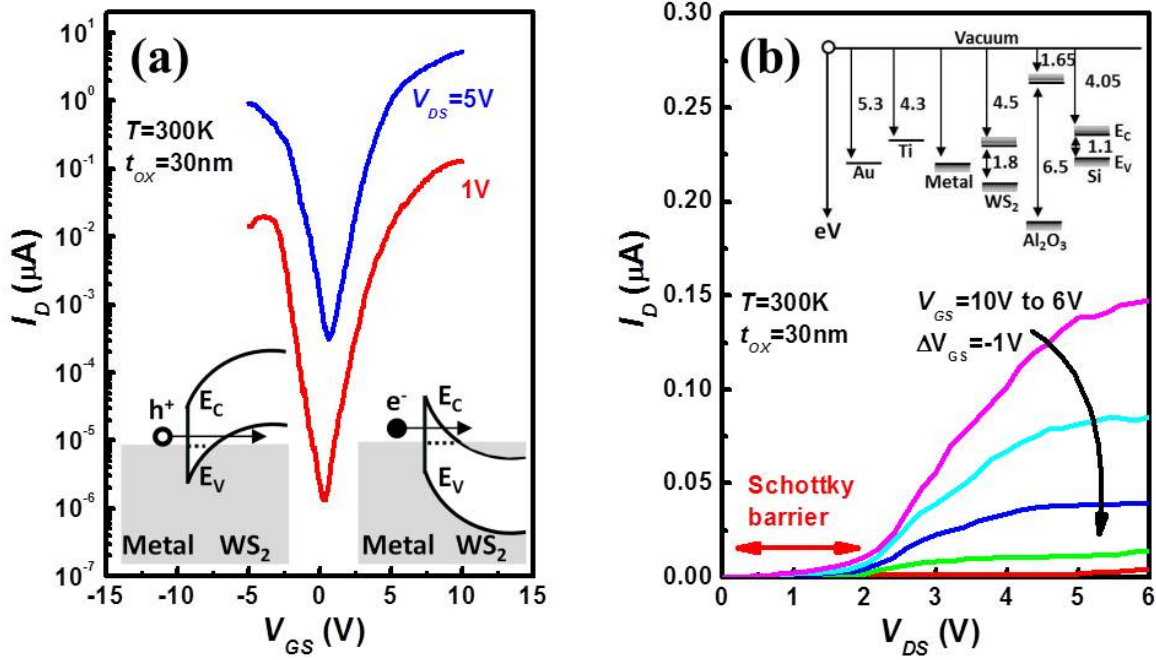

**FIG**. 2. (a) Drain current $I_D$ vs. back gate voltage $V_{GS}$ with $W/L$=2.0/2.5 μm at various drain voltages $V_{DS}$, showing ~$10^5$x on/off current ratio and ambipolar behavior. The insert sketch image shows a conduction mechanism of electrons or holes depending on the gate bias. (b) Drain current $I_D$ vs. drain voltage $V_{DS}$ indicating the presence of Schottky barrier limited-current and current saturation. The band diagram of insert image indicates the formation of Schottky barrier contact between metal and $WS_2$ [26~28].

The optical properties of $WS_2$ nanotubes have been investigated earlier and response to visible light has been reported [29]. In the $WS_2$ flakes studied here, the photoresponse of the WS$_2$ Schottky barrier FETs was measured by illuminating the device with a halogen lamp; the result is shown in Fig. 3(a). The insert image of Fig. 3(a) shows a schematic representation of electron-hole pair generation upon photon absorption, and the increase of drain current due to conduction by these excess carriers. We observe that the saturation drain voltage increases under illumination, which can be attributed to the photogeneration of carriers. The increase in carrier density requires a higher drain voltage to achieve pinch-off in the channel near the drain side. Figure 3(b) shows multiple cycles of the transient photocurrent response under monochromatic illumination at two wavelengths, corresponding to photon energies of 2.1 eV (580 nm) and 1.9 eV (650 nm), both of which are above the expected bandgap of monolayer WS$_2$ (~1.8 eV). Typically, multilayer structures develop indirect bandgaps lower than their monolayer constituents. A more careful photocurrent spectrum measurement can reveal the energy bandgap and its nature (direct or indirect) of multilayer WS$_2$, which is suggested as a future work. The measured temporal response in Fig 3(b) is fast for various applications, but further time-domain studies along these lines can enable the extraction of the diffusion constants of the photogenerated carriers. We also ensure that the photocurrent is caused by WS$_2$ and not the Si substrate, since once the illumination beam is focused outside the WS$_2$ channel region, the photocurrent effect disappears. The gate leakage of the device is lower than drain current by more than 3 orders of magnitude in this experiment.

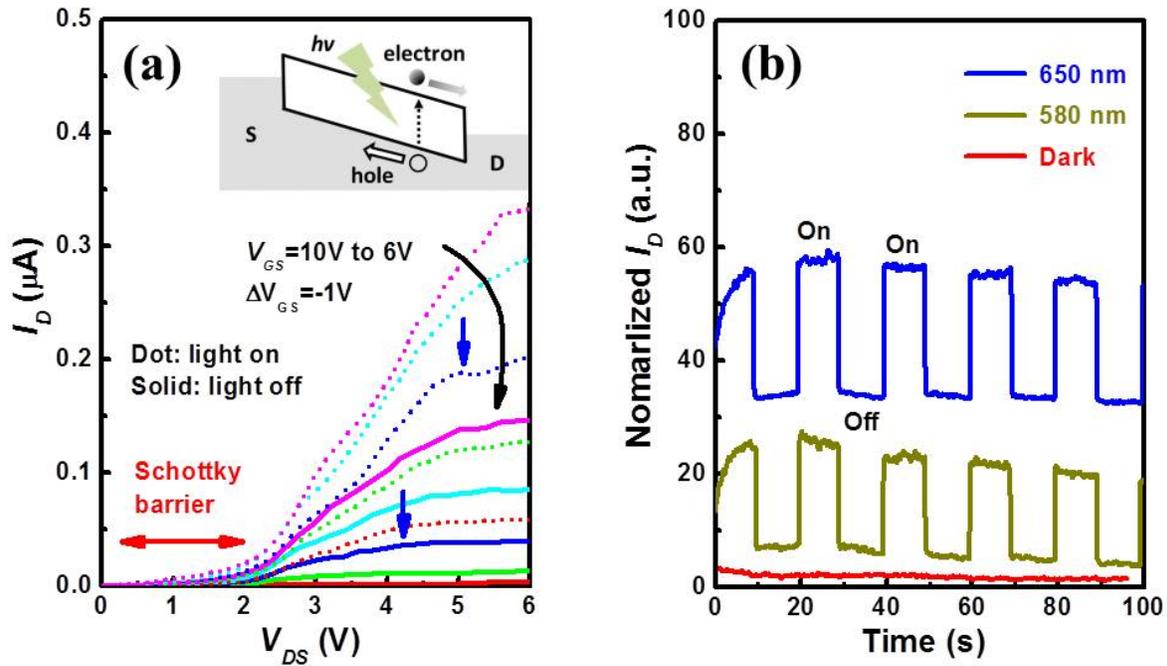

**FIG**. 3. (a) Dependence of Drain current $I_D$ vs. drain voltage $V_{DS}$ at various gate voltage $V_{GS}$ on illumination showing the photocurrent in $WS_2$ FETs. (b) The temporal photocurrent response of 2D $WS_2$ device at $V_{DS}$= 5 V and $V_{GS}$= 0 V at two wavelengths.

In summary, two-dimensional (2D) $WS_2$ transistors were fabricated and characterized for the first time from chemically-synthesized material. Raman measurements prove the 2D crystal nature of the material, and the presence of a bandgap leads to high on/off current ratios and current saturation in the transistors at room temperature. In addition, the observed photoresponse of the 2D layered semiconductor can enable optical device applications. This initial report on the behavior of chemically synthesized $WS_2$ expands the family of 2D crystal semiconductors that possess the features for enabling a wide range of electronic and optical applications. A large amount of work is necessary to form low resistance ohmic contacts and to control the thickness and uniformity of the WS$_2$; the initial observations presented here provide sufficient motivation to move in that direction.

**ACKNOWLEDGEMENTS**

This work was supported by the Semiconductor Research Corporation (SRC), Nanoelectronics Research Initiative (NRI) and the National Institute of Standards and Technology (NIST) through the Midwest Institute for Nanoelectronics Discovery (MIND), the Office of Naval Research (ONR), and the National Science Foundation (NSF). We thank J. Jelenc for technical help in crystal growth, Slovenian Research Agency of the Republic of Slovenia for financial support, contract no. J1-2352 and the Centre of Excellence NAMASTE.